\documentclass[letterpaper,arxiv]{dae-handout}
\graphicspath{{./}}

\newcommand{\trversion}{claude-v3.0}
\newcommand{\trfilename}{Synchronized-Time-Fiction.tex}
\newcommand{\trdate}{02026-MAR-12}
\newcommand{\trshorttitle}{Synchronized Time is a Fiction}
\newcommand{\trauthor}{Paul Borrill}
\newcommand{\traffiliation}{D\AE D\AE LUS}

\usepackage[T1]{fontenc}
\usepackage[utf8]{inputenc}
\IfFileExists{mathdesign.sty}{%
  \usepackage[bitstream-charter]{mathdesign}%
  \DeclareSymbolFont{operators}   {OT1}{cmr} {m}{n}
  \DeclareSymbolFont{letters}     {OML}{cmm} {m}{it}
  \DeclareSymbolFont{symbols}     {OMS}{cmsy}{m}{n}
  \DeclareSymbolFont{largesymbols}{OMX}{cmex}{m}{n}
  \SetSymbolFont{operators}{bold} {OT1}{cmr} {bx}{n}
  \SetSymbolFont{letters}  {bold} {OML}{cmm} {b}{it}
  \SetSymbolFont{symbols}  {bold} {OMS}{cmsy}{b}{n}
  \SetMathAlphabet{\mathit} {normal}{OT1}{cmr}{m}{it}
  \SetMathAlphabet{\mathbf} {normal}{OT1}{cmr}{bx}{n}
  \SetMathAlphabet{\mathsf} {normal}{OT1}{cmss}{m}{n}
  \SetMathAlphabet{\mathtt} {normal}{OT1}{cmtt}{m}{n}
}{%
  \usepackage{charter}%
}

\usepackage{amsmath}
\usepackage{amsthm}
\usepackage{booktabs}
\usepackage{enumitem}
\usepackage{fancyhdr}
\usepackage{titletoc}
\usepackage{etoc}
\usepackage{lastpage}
\usepackage{xcolor}
\usepackage{tikz}
\usepackage{eso-pic}
\usepackage{url}
\usepackage{morefloats}
\usepackage{placeins}
\usepackage{subfiles}
\usetikzlibrary{positioning,shapes.geometric}

\setcitestyle{numbers,square}

\newtheorem{theorem}{Theorem}[section]

\newtheorem{definition}[theorem]{Definition}

\renewcommand{\doi}[1]{\href{https://doi.org/#1}{\nolinkurl{doi:#1}}}

\newcommand{\fito}{\textsc{fito}}

\tikzset{
  badge/.style={
    circle,
    draw=#1,
    fill=#1!10,
    line width=0.5pt,
    minimum size=0.45in,
    font=\tiny\sffamily,
    align=center,
    text=#1!80!black
  }
}

\newcommand{\placebadges}{%
  \ifarxiv\else
  \AddToShipoutPictureBG*{%
    \AtPageUpperLeft{%
      \raisebox{-0.5in}{\hspace{\dimexpr\paperwidth-0.7in\relax}%
        \begin{tikzpicture}[overlay]
          \node[inner sep=0pt] (logo)
            {\includegraphics[height=0.855in]{DAE-Logo.png}};
          \node[badge=green!60!black, left=0.08in of logo]  (b1) {Artifacts\\Available};
          \node[badge=purple!70!black, left=0.08in of b1]   (b2) {Expert\\Verified};
          \node[badge=green!50!black,  left=0.08in of b2]   (b3) {AI\\Assisted};
          \node[badge=blue!70!black,   left=0.08in of b3]   (b4) {Human\\Conceived};
        \end{tikzpicture}%
      }%
    }%
  }%
  \fi
}

\newcommand{\maketrcover}{%
  \ifarxiv\else
  \thispagestyle{empty}
  \begin{fullwidth}
  \vspace*{2in}
  \begin{center}
    {\Large\sffamily\bfseries D\AE D\AE LUS Technical Report}\\[1.5em]
    {\LARGE\sffamily\bfseries Why Synchronized Time is a Fiction:\\[0.3em]
     Daylight Saving Time, Leap Seconds,\\[0.2em]
     and the Guillotine Sharpened for Nothing}\\[2em]
    {\large \trauthor\,,\;\traffiliation}\\[1em]
    {\normalsize \trversion\quad---\quad\trdate}
  \end{center}

  \vspace{2em}
  \noindent\rule{\linewidth}{0.4pt}

  \vspace{1em}
  \footnotesize
  \begin{description}[leftmargin=1.2in, style=sameline, font=\normalfont\scshape]
    \item[Status:]       claude-v3.0 --- Lewis \& Barnes integration
    \item[Filename:]     \texttt{\trfilename}
    \item[Keywords:]     synchronized time, daylight saving time, leap seconds,
                         relativity, FITO, category mistake, causal order,
                         distributed systems, GPS, IEEE~1588
    \item[Related:]      Lamport's Arrow of Time (arXiv, 2026),
                         Cislunar Time (arXiv, 2025),
                         Engineered Simultaneity (arXiv, 2024)
    \item[License:]      \textcopyright\ 2026 \trauthor, \traffiliation.
                         All rights reserved.
  \end{description}

  \vspace{1.5em}
  \noindent\rule{\linewidth}{0.4pt}

  \vspace{2em}
  \begin{center}
    \normalsize\itshape
    This cover page may be discarded when printing.\\
    The paper begins on the following page.
  \end{center}

  \end{fullwidth}
  \clearpage
  \fi
}

\fancypagestyle{plain}{%
  \fancyhf{}%
  \fancyfoot[L]{\small\textsc{\trshorttitle}}%
  \fancyfoot[C]{\small\thepage\ of \pageref{LastPage}}%
  \ifarxiv\else
  \fancyfoot[R]{\scriptsize\texttt{\trfilename}}%
  \fi
  \ifarxiv\else
  \fancyfoot[L]{\raisebox{-0.8in}{\tiny\texttt{\trversion\ -- \trdate}}}%
  \fancyfoot[R]{\raisebox{-0.8in}{\tiny\texttt{\trfilename}}}%
  \fi
}
\title{Why Synchronized Time is a Fiction:\\[0.3em]
Daylight Saving Time, Leap Seconds,\\[0.2em]
and the Guillotine Sharpened for Nothing}
\author[Paul Borrill]{Paul Borrill, D\AE D\AE LUS}
\date{02026-MAR-12}

\begin{document}
\maketrcover
\setcounter{page}{1}
\maketitle
\placebadges
\thispagestyle{plain}

\daemargintoc

\begin{abstract}
\noindent
Civilization maintains an elaborate infrastructure devoted to the maintenance of synchronized time.
Governments mandate daylight saving time.
Standards bodies insert leap seconds into Coordinated Universal Time.
Engineers debate leap milliseconds and leap nanoseconds.
The Global Positioning System applies relativistic corrections at the nanosecond level.
All of these adjustments attempt to preserve an assumption: that a single global time exists and that clocks can be made to agree upon it.

This paper argues that this assumption constitutes a \emph{category mistake} in the sense of Ryle~\citep{ryle1949}.
We show that special and general relativity prohibit absolute simultaneity, that the one-way speed of light is conventionally defined rather than measured, and that recent experiments on indefinite causal order demonstrate nature admits correlations with no well-defined temporal sequence.
We trace the consequences of this category mistake through distributed computing, where it manifests as the \emph{Forward-In-Time-Only} (\fito{}) assumption that underlies Lamport's logical clocks~\citep{lamport1978}, the impossibility results of Fischer--Lynch--Paterson~\citep{flp1985}, and the CAP theorem~\citep{brewer2000}.
From this perspective, daylight saving time and leap seconds are not corrections to time but corrections to conventions---they sharpen the guillotine of synchronization in preparation for executing something that does not exist.
\end{abstract}

\FloatBarrier
\section[Introduction]{Introduction}
\label{sec:intro}

Twice a year, billions of people adjust their clocks by one hour.
Nothing in nature changes when this happens.%
\marginalia{The EU voted in 2019 to abolish DST changes; the measure has been stalled in the Council of Ministers ever since. The United States passed the Sunshine Protection Act in the Senate in 2022 but it failed in the House. Both attempts illustrate how deeply embedded the convention is.}
The Earth does not suddenly rotate faster or slower.
Atomic clocks do not alter their oscillation frequency.
Satellites do not adjust their orbits.
Only the labels on clocks change.

The practice of daylight saving time appears trivial, but it conceals a deeper assumption: that somewhere behind the convention there exists a true time, a global reference that all clocks are supposed to approximate.
Leap seconds express the same assumption at higher precision---they adjust Coordinated Universal Time (UTC) to keep it within one second of the Earth's rotation.
The Global Positioning System applies relativistic corrections to keep its atomic clocks synchronized to a common reference.
Engineers designing high-frequency trading systems, data center networks, and distributed databases devote enormous effort to reducing clock skew, as if the goal of perfect synchronization were physically achievable.

This paper argues that it is not.
The assumption of a global synchronized time is a \emph{category mistake}---a term introduced by Ryle~\citep{ryle1949} to describe the error of treating a concept from one logical category as though it belongs to another.%
\marginalia[-1cm]{Ryle's original example: a visitor to Oxford sees the colleges, libraries, and playing fields, then asks ``But where is the University?'' The university is not a separate entity in the same category as the buildings---it is the organizational pattern \emph{among} them. Global time is the analogous phantom.}
Global time is not a physical quantity that clocks fail to measure accurately.
It is a convention that has been mistaken for a measurement.

The purpose of this paper is to trace this category mistake from its most visible manifestation (daylight saving time) through its scientific foundations (Einstein synchronization, the relativity of simultaneity, Bell's theorem, indefinite causal order) to its engineering consequences (Lamport clocks, the \fito{} assumption, the impossibility results of distributed computing).
We argue that once the category mistake is made explicit, a more coherent alternative emerges: systems designed around causal semantics rather than temporal coordination.

\FloatBarrier
\section[Three Definitions of Time]{Three Competing Definitions of Time}
\label{sec:three-times}

The civilizational infrastructure of timekeeping attempts to reconcile three quantities that do not naturally agree.%
\marginalia{This three-way tension is sometimes called the ``timekeeping trilemma.'' No two of the three can be made to agree without introducing corrections to the third.}

\textbf{Astronomical time} is tied to the Earth's rotation.
One day is the interval between successive solar transits.
This definition served humanity for millennia, but the Earth's rotation is irregular: tidal interactions with the Moon, post-glacial rebound, and unpredictable geophysical processes cause the length of the day to vary by milliseconds.
The standard designation for this quantity is UT1.

\textbf{Atomic time} is defined by the cesium-133 hyperfine transition: exactly 9,192,631,770 oscillations constitute one SI second.
International Atomic Time (TAI) counts these seconds monotonically from an epoch in 1958.
TAI is extraordinarily stable---modern optical lattice clocks achieve fractional uncertainties below $10^{-18}$---but it is completely indifferent to the Earth's rotation.

\textbf{Civil time} is what appears on the clocks that govern human activity: work schedules, train timetables, financial market opening bells.
UTC attempts to bridge the gap between TAI and UT1 by inserting leap seconds.
Daylight saving time applies an additional offset for social convenience.

The critical observation is that these three quantities are not three approximations of the same underlying reality.
They are three different constructions, each defined by different physical or social processes, forced into alignment by convention.
The corrections applied to maintain alignment---leap seconds, DST offsets, relativistic GPS adjustments---are not measurements of error.
They are \emph{negotiations among conventions}.

\FloatBarrier
\section[Leap Seconds]{Leap Seconds: Correcting a Convention}
\label{sec:leap-seconds}

Leap seconds were introduced in 1972 to maintain the constraint
\begin{equation}
\label{eq:leap-constraint}
|t_{\mathrm{UTC}} - t_{\mathrm{UT1}}| < 0.9 \;\text{seconds.}
\end{equation}
When the difference threatens to exceed this bound, the International Earth Rotation and Reference Systems Service (IERS) announces an additional second:
\[
23\!:\!59\!:\!59 \;\rightarrow\; 23\!:\!59\!:\!60 \;\rightarrow\; 00\!:\!00\!:\!00
\]

\marginalia{Since 1972, 27 leap seconds have been inserted. All have been positive (adding a second). The Earth has been slowing overall, but the rate is unpredictable---the decision to insert a leap second cannot be made more than six months in advance.}

The engineering consequences of leap seconds are well documented.
The 2012 leap second caused widespread Linux kernel failures because the kernel's timekeeping code was not designed for a 61-second minute.
Google's response was to implement a ``leap smear''---distributing the extra second across a 24-hour window by slightly adjusting the clock rate.%
\marginalia{Google's leap smear is itself an admission that discrete corrections to a continuous quantity are architecturally problematic. The smear trades accuracy for continuity---precisely the tradeoff that would be unnecessary if the quantity being corrected were physical rather than conventional.}
Amazon, Meta, and Microsoft adopted similar strategies.
None of these approaches measure time more accurately.
They smooth the discontinuity in a convention.

In 2022, the General Conference on Weights and Measures (CGPM) voted to abolish leap seconds by 2035~\citep{itu2022}.
The decision was framed as a practical engineering concession, but it is more revealing than that: it is an institutional admission that the constraint in Equation~\ref{eq:leap-constraint} serves no physical purpose.
Atomic time and astronomical time are simply different things.

\FloatBarrier
\section[Daylight Saving Time]{Daylight Saving Time: The Convention Made Explicit}
\label{sec:dst}

If leap seconds are a subtle correction, daylight saving time is a blunt one.
When DST begins, clocks shift by exactly one hour:
\begin{equation}
\label{eq:dst}
t' = t + 3600 \;\text{seconds.}
\end{equation}
When it ends, the shift is reversed.
No physical process corresponds to this change.
It exists to adjust social schedules relative to daylight.%
\marginalia{The original justification for DST---energy savings---has been repeatedly debunked. A 2008 study by the U.S. Department of Energy found that the 2007 DST extension saved approximately 0.03\% of total electricity consumption. The primary effect of DST is to shift the pattern of human activity, not to save energy.}

The reason DST matters for this paper is that it makes the conventional nature of civil time impossible to deny.
No physicist would argue that the physical universe shifts by one hour on the second Sunday of March.
Yet the same civilizational infrastructure that treats DST as a convention treats UTC as something closer to a physical measurement.
The distinction is one of degree, not of kind.

Both DST and UTC are conventions imposed on an underlying physical reality that does not contain a single global time.
DST simply applies a larger and more visible correction, making the category mistake harder to ignore.

\FloatBarrier
\section[Einstein Synchronization]{Einstein Synchronization and the One-Way Speed of Light}
\label{sec:einstein}

The category mistake in synchronized time has its deepest roots in the physics of simultaneity.
In 1905, Einstein showed that the synchronization of distant clocks requires a \emph{convention}~\citep{einstein1905}.%
\marginalia{Einstein's original paper does not call this a convention---he calls it a ``definition'' (\emph{Festsetzung}). The conventionality was made fully explicit by Reichenbach in 1928 and by Gr\"unbaum in 1963. The point is that the one-way speed of light is \emph{defined}, not measured.}

Consider two clocks $A$ and $B$ separated by distance $d$.
A light signal is sent from $A$ at time $t_1$ (as read by $A$), arrives at $B$, is immediately reflected, and returns to $A$ at time $t_3$.
Einstein's synchronization convention defines the time of arrival at $B$ as
\begin{equation}
\label{eq:einstein-sync}
t_B = t_1 + \frac{1}{2}(t_3 - t_1)
\end{equation}
which is equivalent to assuming that the one-way speed of light is equal in both directions.

The \emph{round-trip} speed of light is measurable: it requires only one clock.
The \emph{one-way} speed of light cannot be independently measured without presupposing a synchronization convention---the very thing it is supposed to establish.%
\marginalia[-0.5cm]{This circularity is not a minor technical point. It means that every clock synchronization protocol---NTP, PTP (IEEE~1588), GPS---inherits a conventional component that cannot be eliminated by better engineering. The convention is built into the physics, not into the apparatus.}
Reichenbach parameterized this freedom with a factor $\varepsilon \in (0,1)$, where Einstein's convention corresponds to $\varepsilon = \frac{1}{2}$:
\begin{equation}
\label{eq:reichenbach}
t_B = t_1 + \varepsilon(t_3 - t_1), \quad 0 < \varepsilon < 1.
\end{equation}
Any value of $\varepsilon$ in the open interval $(0,1)$ is consistent with all observable phenomena.
The choice of $\varepsilon = \frac{1}{2}$ is natural and convenient, but it is not determined by experiment.

This is not a historical curiosity.
It is a theorem about the structure of spacetime, and it constrains every engineering protocol that claims to synchronize distant clocks.
The Stanford Encyclopedia of Philosophy summarizes the situation precisely: all schemes for establishing convention-free synchrony must fail, because any proposed synchronization procedure can itself be described in terms of a nonstandard synchrony.%
\marginalia{Janis (1983) and Norton (1986) proved that no operational procedure can establish synchrony without presupposing it. Winnie (1970) reformulated all of special relativity in terms of nonstandard synchronies, demonstrating that the physics is invariant under the choice of $\varepsilon$. Yet, as the SEP notes, editors of respected journals continue to accept papers purporting to measure one-way light speeds.}
The circularity is not an engineering problem to be solved.
It is a structural feature of the theory.

\subsection{Lewis \& Barnes: The Cosmological Proof}


The conventionality of one-way light speed is not merely a philosophical curiosity.
Lewis and Barnes~\citep{lewis2021} demonstrated that it has cosmological consequences---or, more precisely, that it has no cosmological consequences, which is the whole point.%
\marginalia[1cm]{Lewis was the physicist who advised Derek Muller (Veritasium) on the widely viewed video ``Why No One Has Measured the Speed of Light''~\citep{veritasium2020}. The paper grew from that consultation.}

Following the formalism of Anderson et al.~\citep{anderson1998}, Lewis and Barnes parameterize the one-way speeds of light as
\begin{equation}
\label{eq:lewis-kappa}
c_{\pm} = \frac{c}{1 \mp \kappa}
\end{equation}
where $\kappa = 0$ recovers Einstein's isotropic convention and $\kappa = 1$ gives the extreme anisotropic case: infinite speed in one direction, $c/2$ in the other.
The round-trip speed is always $c$, regardless of $\kappa$.

\begin{figure}[h]
\centering
\includegraphics[width=\textwidth]{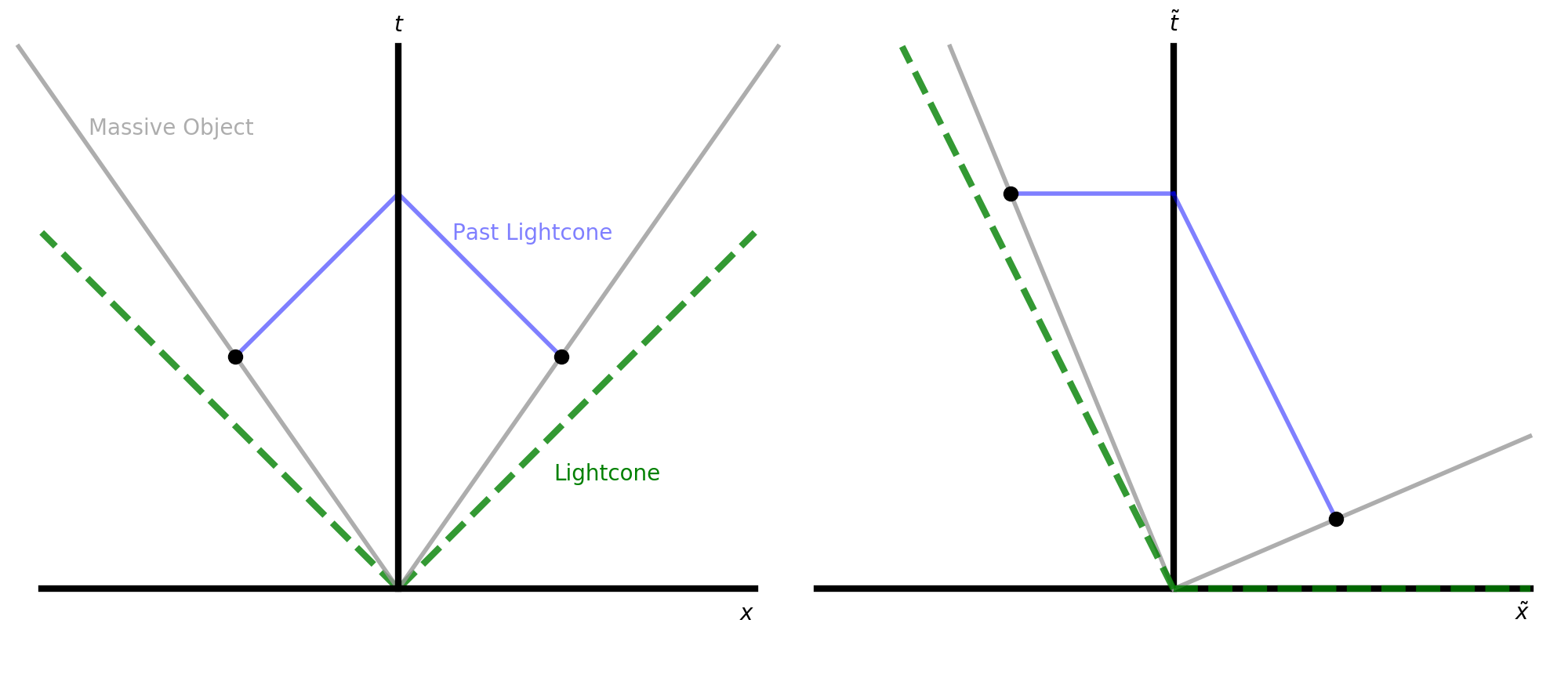}
\caption{Space-time diagrams for isotropic (left) and extreme anisotropic (right) one-way speed of light. Green dashed lines: future light cones. Grey lines: worldlines of massive objects. Blue lines: light emitted after equal proper time. In both cases, the observer at the origin sees the same universe. From Lewis \& Barnes~\citep{lewis2021}, Figure~1.}
\label{fig:lewis-lightcones}
\end{figure}

The key result is this: in the Milne universe---the empty, flat limiting case of the Friedmann-Robertson-Walker metric---an anisotropic speed of light produces \emph{anisotropic time dilation} that exactly compensates for the differing light travel times.
The modified Lorentz factor becomes
\begin{equation}
\label{eq:lewis-gamma}
\tilde{\gamma} = \frac{1 - \kappa v}{\sqrt{1 - v^2}}
\end{equation}
and the resulting view of the cosmos is isotropic for \emph{every} value of $\kappa$.
The universe looks the same whether the speed of light is equal in both directions or infinite in one direction and $c/2$ in the other.%
\marginalia{The anisotropic-speed advocate must conclude that galaxies at a given distance have different recession speeds in different directions, and that the universe is expanding faster to the left than to the right. But the dependence of redshift on the speed of light means that none of this is observable. The initial velocities have exactly the right anisotropy to balance the anisotropic redshift. There is, as Lewis and Barnes note, nothing internally inconsistent about this.}

This is not merely a change of coordinates.
It is a proof, from special relativity alone, that the one-way speed of light is \emph{not an observable}.
No measurement---terrestrial or cosmological---can distinguish $\kappa = 0$ from $\kappa = 1$.
The quantity is conventionally defined, exactly as Reichenbach's $\varepsilon$ parameter indicates.

\subsection{The Photon's Perspective}


There is a way to see why this must be so that does not require the full apparatus of coordinate transformations.
Consider what happens when a photon leaves one system for the first time.
Velocity is not defined, because both distance and time are not definable.
Photon proper time is always zero, no matter how far the photon travels.
Until a photon arrives, distance does not exist.%
\marginalia{This is the ``Life of a Photon'' perspective. A photon arriving at the JWST has traversed the largest distances in the known universe, but it has done so in zero proper time. It carries no timestamps. There are no Minkowski manifold coordinates transmitted alongside the photon. All we have is the arrival of information: wavelength, polarization, and phase.}

The first observer of that photon is also unable to determine the distance between the emitter and itself.
However, if that first receiver acts as a mirror and reflects the photon back to the original emitter, then it becomes possible---in principle---for the original emitter to determine the distance of the two-way traversal, by referencing some local mechanism that can do some counting on its behalf.
As with all quantum clocks, you need one clock in order to measure another.

This lack of definability of distance in the one-way traversal, combined with the photon's proper time being zero, provides a physical intuition for why we need a round trip to fully specify information transfer, and why one-way traversals are insufficient.
It also provides an intuition for why we achieve correlations in Bell state measurements while still requiring conventional speed-of-light transfers for signaling: this corresponds precisely to the Lewis and Barnes extreme case, where the speed is infinite in one direction and $c/2$ in the other.

\subsection{Indefinite, Not Infinite}


Instead of adopting a Newtonian infinite speed of light for the inbound photon, can we obtain an equivalent result by recognizing the initial one-way traversal as \emph{indefinite}?
The Lewis and Barnes bounds include both zero and infinity in the extreme anisotropic case (Equation~\ref{eq:lewis-kappa}).
This perspective on spacetime being indefinite appears consistent with laboratory measurements that have decisively demonstrated---to almost seven standard deviations---the indefinite arrival order of a quantum witness in the quantum switch configuration~\citep{rubino2017}.%
\marginalia[-3cm]{The connection between the conventionality of one-way light speed (a result from special relativity) and indefinite causal order (a result from quantum mechanics) is suggestive but not yet rigorous. If the one-way speed is not an observable, and causal order can be indefinite, then the entire edifice of timestamp-based event ordering rests on two independent conventions, either of which can be varied without changing any observable.}

The word ``indefinite'' is deliberate.
The one-way speed of light is not infinite, and it is not any particular finite value.
It is \emph{undefined}---a quantity that has no physical meaning independent of a convention.
This is not a limitation of our measurement technology.
It is a feature of the theory.

\begin{figure}[h]
\centering
\includegraphics[width=\textwidth]{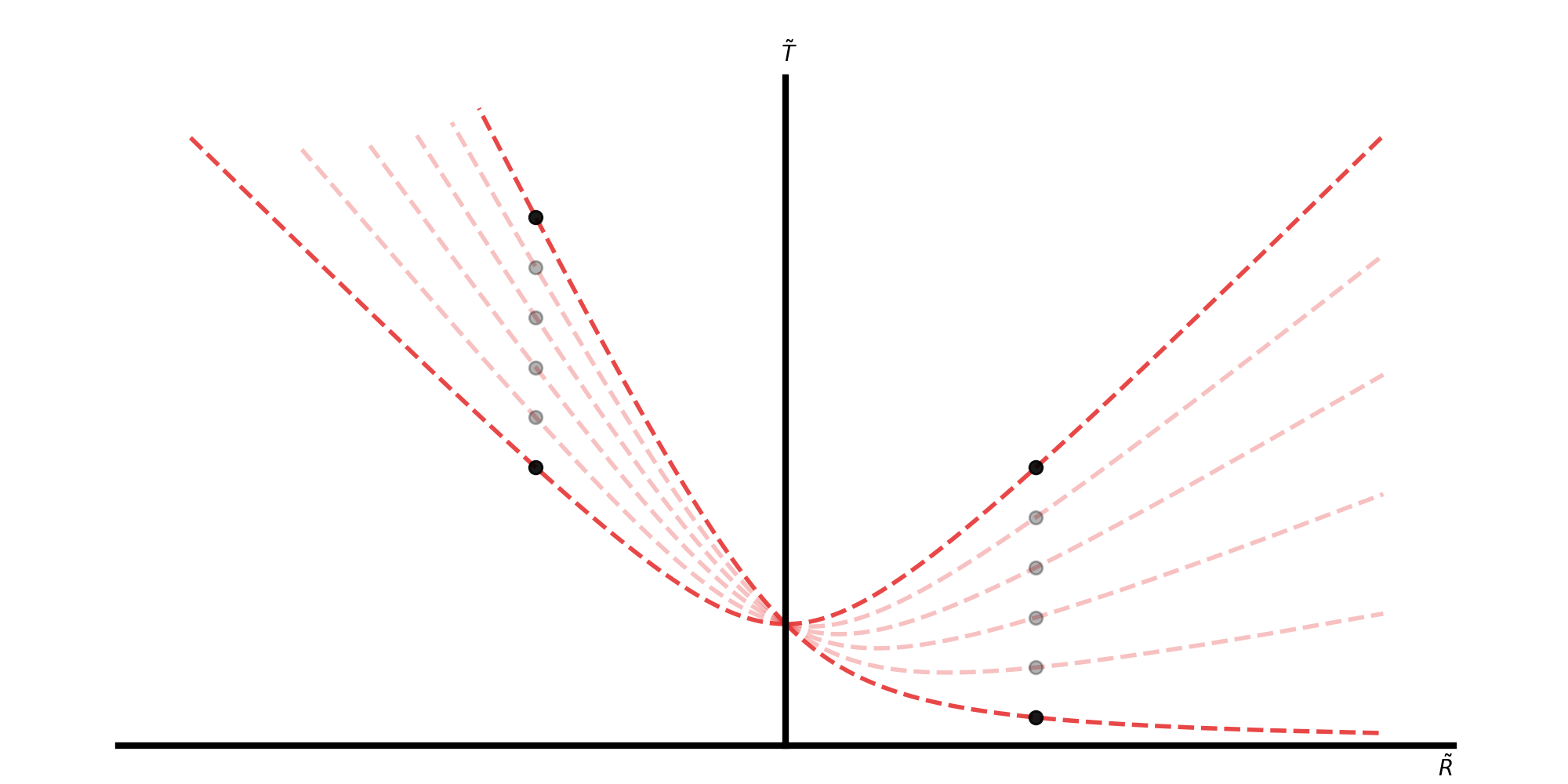}
\caption{Lines of simultaneity in the emitter's FRW coordinates, mapped into anisotropic velocity-of-light coordinates from $\kappa = 0$ (isotropic, hyperbola) to $\kappa = 1$ (extreme anisotropic, bold red), with intermediate cases in steps of $\kappa = 0.2$ (lighter red). The filled circles mark the emitter's location for each case. The spatial position in the $\tilde{R}$ coordinate is independent of $\kappa$---only the simultaneity surface deforms. From Lewis \& Barnes~\citep{lewis2021}, Figure~5.}
\label{fig:lewis-simultaneity}
\end{figure}

\FloatBarrier
\section[Relativity of Simultaneity]{The Relativity of Simultaneity}
\label{sec:simultaneity}

Special relativity does not merely make synchronization difficult.
It makes absolute simultaneity \emph{impossible}.

\begin{figure}[h]
\centering
\includegraphics[width=\textwidth]{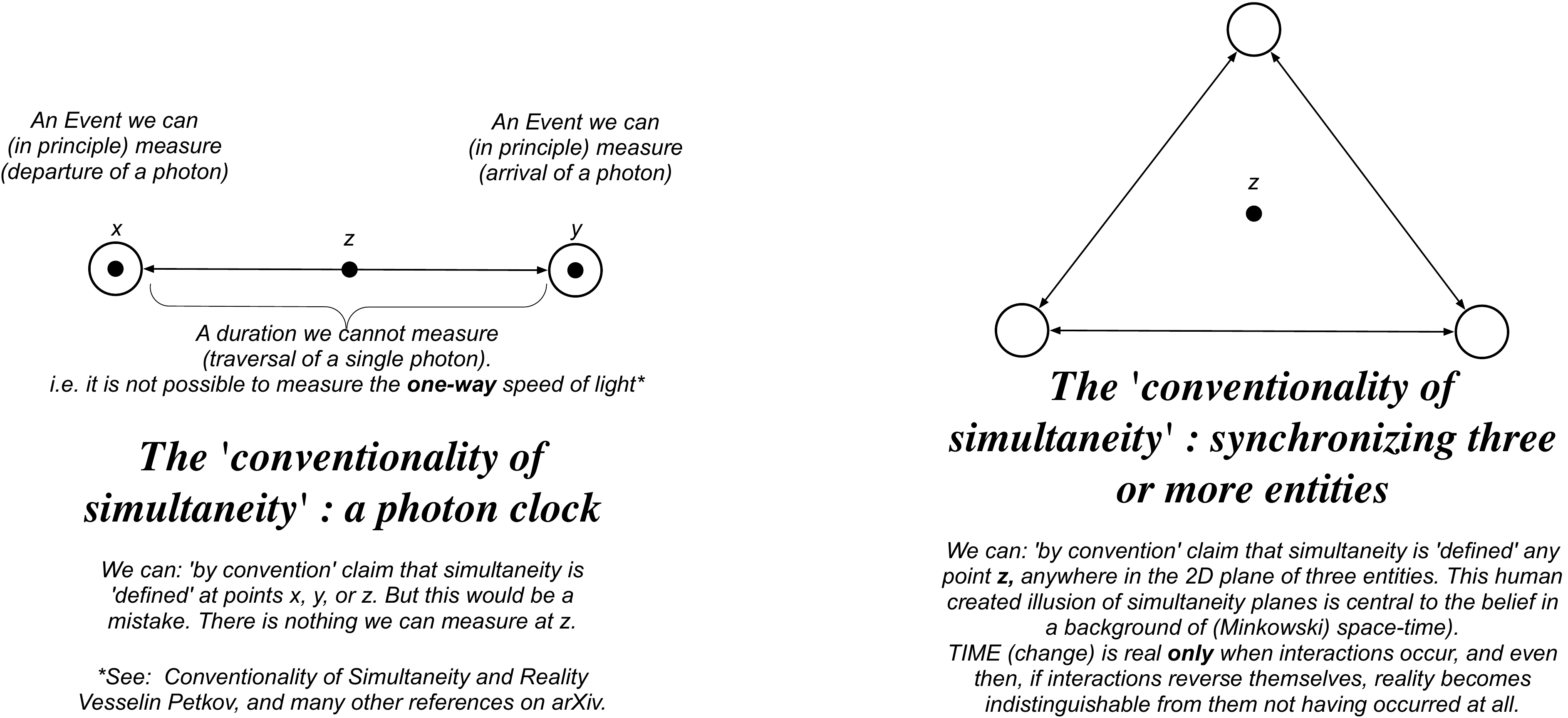}
\caption{The conventionality of simultaneity: for spacelike-separated events, different inertial observers disagree on which event occurred first. The ``hyperplanes of simultaneity'' that the clock synchronization hypothesis presupposes (Definition~\ref{def:clock-sync-hypothesis}) are observer-dependent constructions, not objective features of spacetime.}
\label{fig:conventionality}
\end{figure}

For two events separated by spatial distance $\Delta x$ and time interval $\Delta t$ in one inertial frame, the Lorentz transformation gives the time interval in a frame moving at velocity $v$:
\begin{equation}
\label{eq:lorentz}
\Delta t' = \gamma\!\left(\Delta t - \frac{v\,\Delta x}{c^2}\right), \quad \gamma = \frac{1}{\sqrt{1 - v^2/c^2}}.
\end{equation}
If $\Delta t = 0$ in one frame (the events are simultaneous), then $\Delta t' = -\gamma v \Delta x / c^2$ in the other.
For spacelike-separated events, the sign of $\Delta t'$ depends on $v$: different observers disagree not merely on \emph{when} the events occurred but on \emph{which one occurred first}.%
\marginalia{This is not a practical limitation of measurement technology. It is a structural feature of spacetime geometry. No improvement in clock precision, no refinement of synchronization protocol, can overcome it. The ordering of spacelike-separated events is not a fact about the world---it is a fact about the observer.}

General relativity compounds the problem.
In a gravitational field, clocks at different heights run at different rates.
A clock at sea level ticks approximately 1 part in $10^{16}$ slower per meter of altitude than a clock on a mountaintop~\citep{hafele1972}.
This is not negligible: the GPS system applies a relativistic correction of approximately $+38$ microseconds per day to its satellite clocks to compensate for the combined effects of velocity (special relativity, $-7\;\mu$s/day) and gravitational potential (general relativity, $+45\;\mu$s/day)~\citep{ashby2003}.%
\marginalia{Without the relativistic correction, GPS position errors would accumulate at roughly 10~km per day. GPS works not because its clocks are synchronized to a single global time, but because the relativistic corrections are applied \emph{by convention} to make the satellite clocks behave \emph{as if} they were at sea level on a non-rotating Earth.}

The GPS correction is instructive.
It does not synchronize satellite clocks to a physically real global time.
It adjusts them to a \emph{coordinate time}---a mathematical construction chosen for computational convenience.
The correction is exact only for the adopted model of Earth's gravitational field.
It is a convention, not a measurement.

\FloatBarrier
\section[Clock Synchronization Hypothesis]{The Clock Synchronization Hypothesis and the Stanford Data}
\label{sec:stanford}

The theoretical arguments of the preceding sections are not merely academic.
They make empirical predictions that have already been observed---and misinterpreted.

In 2018, Geng et al.\ at Stanford presented a system for nanosecond-level clock synchronization within a data center, supervised by Mendel Rosenblum.%
\marginalia{Y.~Geng et al., ``Exploiting a Natural Network Effect for Scalable, Fine-grained Clock Synchronization,'' NSDI~2018. The system achieved sub-25-nanosecond synchronization across a datacenter fabric using a network of boundary clocks and pairwise calibration.}
The system achieved remarkable precision.
But the measured data contained an anomaly that the authors characterized as ``bad data'': packets whose \emph{received} timestamps were earlier than their \emph{sent} timestamps.

\begin{figure}[h]
\centering
\includegraphics[width=\textwidth,trim=8mm 5mm 12mm 0cm,clip]{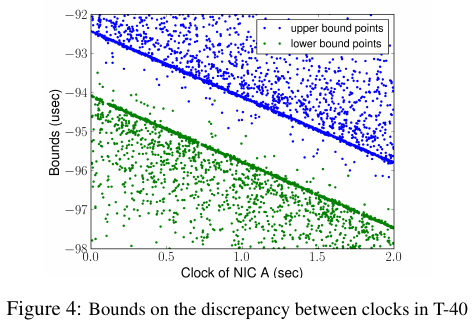}
\caption{Bounds on timestamp discrepancy from the Stanford nanosecond synchronization measurements (reproduced from Geng et al., NSDI 2018, Figure~4). The thick blue and green lines represent the minimum spacetime intervals between sending and receiving timestamps. Points appearing within the ``forbidden zone'' between these bounds were classified as measurement errors. We argue they validate the relativity of simultaneity.\\[0.5em]
Note the vertical axis: the discrepancies are measured in \emph{microseconds} ($\mu$s), yet the Stanford paper claims nanosecond synchronization accuracy---three orders of magnitude smaller than the anomalies being discarded. The ``forbidden zone'' is not at the edge of the measurement noise floor. It is a thousand times larger than the claimed precision.}
\label{fig:stanford-bounds}
\end{figure}

The authors applied a Support Vector Machine classifier to separate ``good'' from ``bad'' data points---a hyperplane-based technique that presupposes the existence of hyperplanes of simultaneity.
This is precisely the assumption that relativity denies.

The data points in the forbidden zone admit an alternative interpretation: they are not measurement errors.
They are empirical evidence that the three implicit assumptions behind the \emph{clock synchronization hypothesis} do not hold.%
\marginalia[-1cm]{The term ``clock synchronization hypothesis'' is our name for the cluster of assumptions that most computer scientists treat as axiomatic. Making these assumptions explicit is the first step toward recognizing the category mistake they embody.}
Those assumptions are:

\begin{definition}[The Clock Synchronization Hypothesis]
\label{def:clock-sync-hypothesis}
The clock synchronization hypothesis assumes that:
\begin{enumerate}
    \item \textbf{Smooth:} There is a smooth, diffeomorphic relationship between all pairs of clocks in the measurement field---i.e., hyperplanes of simultaneity exist.
    \item \textbf{Irreversible:} All clocks count time in one direction only---monotonically increasing, consistent with the second law of thermodynamics.
    \item \textbf{Global:} There is a smooth background connecting all clocks, against which a ``one-way speed of propagation'' can be defined.
\end{enumerate}
\end{definition}

Each of these assumptions fails under scrutiny.
The first (smooth) implies the existence of the very simultaneity planes that special relativity denies.
The second (irreversible) conflates a macroscopic statistical property---entropy increase---with a fundamental law; at the microscopic level, all fundamental physical processes are time-reversible.
The third (global) requires exactly the kind of background time that general relativity replaces with local, observer-dependent time coordinates.%
\marginalia{The third assumption is the most pernicious, because it is also the most invisible. Computer scientists rarely state it explicitly. They simply assume that when they write $t_{\mathrm{recv}} - t_{\mathrm{send}}$, the subtraction is meaningful. It is meaningful only within a chosen synchronization convention.}

The Stanford data is not an isolated case.
Any high-precision timestamp measurement in a distributed system will, at sufficient resolution, produce events that violate the clock synchronization hypothesis.
The standard response is to discard these events as outliers.
The correct response is to recognize them as evidence that the hypothesis is wrong.

The implications are immediate:

Timestamps are an inherently unsafe ordering construct.
An incrementing clock on one computer cannot be presumed valid by another computer and used for control---whether for consistency operations, event ordering, or regulatory compliance.
Applications such as high-frequency trading, where different servers apply timestamps to incoming transactions, will necessarily produce inconsistent orderings for some non-trivial fraction of transactions, even when the system is working perfectly.

\FloatBarrier
\section[Quantum Causal Order]{Bell's Theorem and Indefinite Causal Order}
\label{sec:quantum}

If relativity undermines absolute simultaneity, quantum theory undermines the assumption that causal order is globally well-defined.

Bell's theorem~\citep{bell1964} establishes that no locally causal hidden-variable theory can reproduce the correlations predicted by quantum mechanics.
The CHSH form of the Bell inequality states that for any local hidden-variable model,
\begin{equation}
\label{eq:chsh}
|E(a,b) + E(a,b') + E(a',b) - E(a',b')| \leq 2
\end{equation}
where $E(a,b)$ denotes the expectation value of the product of measurement outcomes for settings $a$ and $b$.
Quantum mechanics predicts---and experiment confirms---violations up to $2\sqrt{2}$.%
\marginalia{The experimental violations of Bell inequalities have been confirmed in loophole-free experiments since 2015. The correlations are real. What they mean for the causal structure of nature is the subject of ongoing debate, but all interpretations agree that classical, locally ordered causation is insufficient.}

More recently, the framework of \emph{indefinite causal order}~\citep{oreshkov2012} has shown that quantum mechanics permits processes in which the causal ordering of operations is not merely unknown but \emph{undefined}.
The process matrix formalism describes such systems:
\begin{equation}
\label{eq:process-matrix}
P(a,b \,|\, x,y) = \mathrm{Tr}\!\left[W \left(A_{a|x} \otimes B_{b|y}\right)\right]
\end{equation}
where $W$ is a process matrix that is not constrained to respect any fixed causal ordering between $A$ and $B$.

Rubino et al.~\citep{rubino2017} experimentally demonstrated a quantum switch---a device that places two operations in a superposition of causal orders.
Proietti et al.~\citep{proietti2019} extended this to show that even the notion of objective observed events may be observer-dependent.

\subsection{The Bell Test Circularity}

The Bell test itself reveals a deeper circularity that connects directly to the synchronization problem.%
\marginalia{This argument was first developed in DAE-E-007 (February 2026) for a conversation with Kevin Stanton on IEEE~1588 and clock synchronization. It applies with equal force here.}

To establish that two measurements in a Bell test are spacelike-separated---meaning neither could have causally influenced the other at light speed---the experimenters must compare the \emph{timestamps} of the two detection events.
This comparison requires synchronized clocks at the two detector stations.
But we have established (Section~\ref{sec:einstein}) that synchronized clocks, in the absolute sense, do not exist.

What the experimenters actually have is a pair of clocks whose disagreement has been modeled and corrected to within some tolerance, using protocols that assume specific things about signal propagation, path symmetry, and the one-way speed of light---the very quantities that cannot be independently measured without presupposing synchronization.

The spacelike separation of the measurement events is therefore not a fact about the events.
It is a fact about the synchronization convention applied to the clocks used to timestamp them.
Change the convention, and the same pair of events could be timelike-separated---meaning one \emph{could} have causally influenced the other through ordinary, local, subluminal processes.

The argument for nonlocality thus has the following logical structure:%
\marginalia[-0.5cm]{This circularity is not a ``loophole'' in the sense that experimentalists use the term---it cannot be closed by better detectors or faster random-number generators. It is a structural feature of the argument itself. You cannot establish nonlocality using tools whose calibration assumes a specific causal structure.}

\begin{enumerate}[leftmargin=1.2cm]
    \item We synchronize two distant clocks using a procedure that assumes the one-way speed of light is isotropic.
    \item Using those clocks, we establish that two events are spacelike-separated.
    \item We observe correlations between those events that cannot be explained by local hidden variables.
    \item We conclude that nature is nonlocal.
\end{enumerate}

\noindent But step~1 already contains an assumption about the causal structure of spacetime.
If that assumption is wrong---or even merely conventional rather than factual---then step~2 is undermined, and step~4 does not follow.

The correlations are real.
The violation of Bell's inequality is genuine.
But the interpretation---that these correlations require nonlocal causation---may be an artifact of the measurement ontology: the insistence on assigning definite, linearly-ordered timestamps to each detection event using classically synchronized clocks.%
\marginalia{This is roughly analogous to projecting a three-dimensional object onto a two-dimensional plane and then declaring that the object has impossible geometry because the projection contains overlapping lines. The impossibility is in the projection, not the object.}

The irony is acute: quantum mechanics is the most accurate physical theory ever constructed, describing a world in which superposition, entanglement, and interference are fundamental.
Yet we measure this profoundly non-classical domain using the most classical apparatus imaginable---a pair of clocks ticking forward in time, synchronized by a procedure that assumes Newtonian-style absolute simultaneity.
We use \fito{} measurement protocols to interrogate a theory whose formalism is time-symmetric.

The implication for synchronized time is fundamental: if nature itself does not always assign a definite temporal order to events, then the civilizational project of making all clocks agree on such an order is not merely impractical.
It is an attempt to enforce a structure that the universe does not possess.

\FloatBarrier
\section[Distributed Systems]{Distributed Systems and the \fito{} Assumption}
\label{sec:distributed}

The category mistake is not confined to physics.
It pervades distributed computing.

Lamport's 1978 paper~\citep{lamport1978} freed distributed systems from dependence on synchronized physical clocks by introducing the happens-before relation: a partial order over events defined by program order and message passing.%
\marginalia{Lamport's paper has over 14,000 citations. Its influence on distributed systems is difficult to overstate. The argument here is not that Lamport was wrong to introduce logical clocks---he was brilliantly right. The argument is that the happens-before relation carries an implicit physical assumption that has gone unexamined.}
Logical clocks embed this partial order into the natural numbers via the update rule
\begin{equation}
\label{eq:lamport-clock}
C_i = \max(C_i, C_j) + 1.
\end{equation}
Vector clocks~\citep{fidge1988,mattern1989} refine the embedding to capture the full partial order.

As we have argued elsewhere~\citep{borrill2026lamport}, these constructions carry three implicit commitments that we call the \emph{Forward-In-Time-Only} (\fito{}) assumption:

\begin{definition}[Forward-In-Time-Only (\fito{})]
\label{def:fito}
A distributed systems model satisfies \fito{} if it assumes:
\begin{enumerate}
    \item \textbf{Temporal monotonicity:} every causal chain is mapped to a strictly increasing sequence.
    \item \textbf{Asymmetric causation:} if event $a$ can influence event $b$, then $b$ cannot influence $a$.
    \item \textbf{Global causal reference:} the partial order is observer-independent.
\end{enumerate}
\end{definition}

\marginalia{Compare with Section~\ref{sec:simultaneity}: two observers in different inertial frames \emph{do not} agree on the temporal ordering of spacelike-separated events. Axiom~3 of \fito{} is therefore stronger than what relativity permits.}

The \fito{} assumption is the distributed-systems analogue of the assumption that leap seconds correct: that there exists a single, globally consistent temporal direction against which all events can be ordered.
It is the same category mistake, expressed in different formalism.

\FloatBarrier
\section[IEEE 1588]{IEEE~1588 and the Precision Illusion}
\label{sec:ieee1588}

The Precision Time Protocol (PTP, IEEE~1588)~\citep{ieee1588} is the engineering apotheosis of the synchronization assumption.
It aims to synchronize clocks across a network to sub-microsecond accuracy by measuring message propagation delays and correcting for asymmetries.

PTP's core mechanism involves exchanging four timestamps ($t_1$, $t_2$, $t_3$, $t_4$) between a designated master and a slave device, then computing:
\begin{equation}
\label{eq:ptp-offset}
\mathrm{offset} = \frac{(t_2 - t_1) - (t_4 - t_3)}{2}
\qquad\qquad
\mathrm{delay} = \frac{(t_2 - t_1) + (t_4 - t_3)}{2}
\end{equation}
The slave clock is then adjusted by the computed offset.%
\marginalia{This is conceptually equivalent to Einstein synchronization (Section~\ref{sec:einstein}). The formula assumes symmetric path delay---the $\varepsilon = \frac{1}{2}$ convention. In a datacenter, this assumption fails whenever traffic is asymmetric, switches buffer differently in each direction, or fiber paths differ in length.}

PTP achieves remarkable precision within the constraints of its assumptions.
But those assumptions inherit every problem identified in this paper:%
\marginalia{IEEE~1588 is the protocol most commonly cited by financial regulators (MiFID~II, SEC Rule~613) as the basis for timestamp accuracy requirements. The regulations mandate sub-microsecond timestamps, implicitly assuming that such timestamps have a unique physical meaning. They do not---they have a \emph{conventional} meaning tied to a \emph{chosen} reference.}

\begin{enumerate}
    \item PTP assumes that the one-way propagation delay is half the round-trip delay (the Einstein $\varepsilon = \frac{1}{2}$ convention, Equation~\ref{eq:reichenbach}).
    \item PTP assumes that the master clock's time is ``correct''---but the master is itself synchronized to a GPS or atomic reference that carries conventional corrections.
    \item PTP assumes that network delay is symmetric; in practice, switch buffering, traffic load, and fiber routing introduce asymmetries that are correctable only statistically.
    \item PTP assumes that higher precision is always better---but at the nanosecond scale, relativistic effects (Section~\ref{sec:simultaneity}) and quantum indeterminacy (Section~\ref{sec:quantum}) dominate.
\end{enumerate}

\subsection{Timestamps Are Contracts, Not Truth}

A timestamp does not describe what happened.
It describes what the system agrees to pretend happened.%
\marginalia{This distinction is critical for high-frequency trading, where timestamps become legally binding facts even when the underlying physical order is ambiguous or unknowable. Measurement uncertainty is converted into enforceable certainty. Governance is disguised as engineering.}

In high-frequency trading, timestamps acquire legal authority through regulations like MiFID~II and SEC Rule~613.
The regulatory framework treats timestamps as neutral engineering measurements---objective records of when events occurred.
But we have seen that the ``when'' depends on a synchronization convention, and that convention inherits an irremovable conventional component from the one-way speed of light.

What is actually being enforced is not temporal ordering but \emph{contractual ordering}: an agreement among parties to treat a particular clock's output as authoritative.
This is not a measurement.
It is a social convention with the force of law---and like daylight saving time, it can be changed by fiat because it was never physical to begin with.

The high-frequency trading industry understands this at some level, even if it does not articulate it.
Profits in HFT are derived not primarily from raw speed but from control over where time is measured, when clocks are sampled, how ties are broken, and whose clocks count.
This is temporal rent-seeking: the exploitation of ambiguity that cannot be independently audited by most market participants.

Each additional digit of precision sharpens the guillotine.
It does not bring the clocks closer to a physically real global time.
It brings them closer to a more precise convention.

\FloatBarrier
\section[One-Way Delay]{The One-Way Delay Metric: An Unmeasurable Quantity}
\label{sec:one-way-delay}

The category mistake in synchronized time is codified most explicitly in RFC~7679, which defines a ``one-way delay'' metric for IP networks.%
\marginalia{G.~Almes, S.~Kalidindi, and M.~Zekauskas, ``A One-Way Delay Metric for IP Performance Metrics (IPPM),'' RFC~7679, IETF, 2016. The earlier RFC~3393 defines ``IP Packet Delay Variation'' (PDV) and inherits the same assumptions.}
The metric is defined as:
\[
\text{One-Way Delay} = T_{\mathrm{arrival}}^{B} - T_{\mathrm{departure}}^{A}
\]
This subtraction is only meaningful if the clocks at $A$ and $B$ are synchronized.
But we have established that synchronization requires a convention about the one-way speed of light (Section~\ref{sec:einstein}), that no operational procedure can establish this convention without presupposing it, and that the resulting ``synchronization'' is frame-dependent (Section~\ref{sec:simultaneity}).

One-way delay is therefore not an observable quantity in the relativistic sense.
It is a derived, model-dependent metric---a number that emerges from a chain of conventions, not from a measurement.

The round-trip time (RTT) is measurable: it requires only one clock.
Any metric derived from subtracting timestamps at two different locations inherits the full weight of the synchronization problem.
RFC~7679 acknowledges the need for synchronized clocks but treats the synchronization error as a practical limitation rather than a fundamental one.
This is the category mistake in protocol form: treating a conventional quantity as an approximation to a physical one.

The same critique applies to RFC~3393 (IP Packet Delay Variation), which attempts to mitigate the synchronization problem by expressing delay variation as a difference of two one-way delays:
\[
\text{PDV} = |(t_2 - t_1) - (t_4 - t_3)|
\]
where $t_1$, $t_3$ are send times and $t_2$, $t_4$ are receive times.
This implicitly assumes that clock drift is linear and slow, and that delay variation dominates over clock noise---assumptions that break down at the nanosecond precision that modern networks demand.

\FloatBarrier
\section[Sharpened Guillotine]{The Sharpened Guillotine}
\label{sec:guillotine}

The pattern is now clear.
At every scale---from the one-hour DST offset to the nanosecond GPS correction to the sub-nanosecond PTP synchronization target---the infrastructure of timekeeping applies corrections to maintain agreement among clocks.
Each correction assumes that a true time exists and that clocks are failing to measure it accurately.

We propose that this assumption is backwards.
The clocks are not failing.
There is no global time for them to fail at measuring.

Proposals for leap milliseconds, leap microseconds, and leap nanoseconds~\citep{barbour1999} represent the limiting case of this logic: ever-finer corrections to an ever-more-precise convention.
They sharpen the guillotine---the apparatus of synchronization---in preparation for an execution that will never occur.%
\marginalia{The metaphor is precise: a guillotine that has been sharpened to atomic perfection is still useless if there is nothing to execute. The ``something'' that synchronized time proposes to capture---a single global temporal reference---does not exist in the physics.}

\FloatBarrier
\section[Causal Semantics]{Toward Causal Semantics}
\label{sec:causal}

If global synchronized time is a fiction, what replaces it?

The answer that emerges from both physics and distributed computing is the same: \emph{causal relationships}.
Instead of asking ``when did this event occur?'' we ask ``what did this event depend on, and what depends on it?''

In distributed systems, this shift has been partially realized through vector clocks, causal broadcast protocols, and conflict-free replicated data types (CRDTs).
But these constructions still operate within the \fito{} framework---they assume a globally consistent partial order over events.%
\marginalia{CRDTs achieve eventual consistency by ensuring that concurrent operations commute. This is a significant advance over timestamp-based coordination, but it still presupposes that ``concurrent'' is an observer-independent property. Relativity says it is not.}

A more fundamental approach, which we have developed in a series of companion papers~\citep{borrill2024es, borrill2026lamport, borrill2026flp, borrill2026cap}, is to replace the temporal partial order with an information-theoretic primitive: \emph{mutual information conservation}.
Under this framework, the fundamental invariant is not that events are ordered in time but that information exchanged between processes is conserved---that no information is created or destroyed by the communication itself.

This is the engineering expression of a principle that physics has recognized for over a century: that the laws of nature are time-symmetric, and that the apparent arrow of time is an emergent, statistical phenomenon rather than a fundamental constraint~\citep{barbour1999}.
Daylight saving time, leap seconds, GPS corrections, and IEEE~1588 synchronization are all symptoms of an engineering tradition that has not yet absorbed this insight.

\FloatBarrier
\section[Conclusion]{Conclusion}
\label{sec:conclusion}

Daylight saving time appears trivial.
It is not.
It is the most visible symptom of a category mistake that pervades timekeeping from the civilizational to the quantum scale.

The mistake is to treat global synchronized time as a physical quantity rather than a convention.
This mistake leads to leap seconds that attempt to reconcile two fundamentally different physical processes, GPS corrections that enforce a coordinate time chosen for computational convenience, IEEE~1588 protocols that sharpen synchronization to precisions where the concept itself dissolves, and distributed systems models that embed an observer-independent arrow of time as an unexamined axiom.

Physics offers no single global time.
Relativity makes simultaneity observer-dependent.
Quantum theory makes causal order potentially indefinite.
The engineering response should not be to pursue ever-more-precise synchronization---to sharpen the guillotine---but to design systems around the causal structure that nature actually provides.

The blade is sharp enough.
There is nothing to execute.

\vspace{1em}
{\small\noindent\textbf{Acknowledgments.}
The author thanks Geraint F.~Lewis (University of Sydney) for illuminating
discussions on the conventionality of simultaneity and the one-way speed
of light; Ahmad Byagowi (Meta) and Kevin Stanton (Intel) for detailed
technical engagement with the clock synchronization arguments; and
the Mulligan Stew group for weekly discussions that sharpened the
category mistake framework.
AI tools (Claude, Anthropic) assisted with literature review and
drafting.}

\vspace{1em}

%
%
%
%
%


\end{document}